\documentclass[twocolumn,pra,aps,showpacs]{revtex4}

\usepackage{psfrag,graphicx}
\usepackage{dcolumn}
\usepackage{amsmath,amssymb}
\usepackage{bm}
\usepackage[dvips]{color}

\usepackage{overpic}

\definecolor{mygrey}{gray}{0.35}
\definecolor{mygreen}{rgb}{0.85,1,0.9}
\definecolor{myzard}{cmyk}{0,0,0.05,0}
\definecolor{mywhite}{rgb}{1,1,1}
\definecolor{myred}{rgb}{1,0,0}

 \def\ee{\mathord{\rm e}}
 
 \def\ii{\mathord{\rm i}}

\def\half{\textstyle\frac{1}{2}}

\renewcommand{\ii}{{\rm i}}
\renewcommand{\ee}{{\rm e}}

 \newcommand{\ket}[1]{|#1\rangle}
 \newcommand{\bra}[1]{\langle #1|}

\begin{document}

\title[Short Title]{Non-relativistic limit in the 2+1 Dirac Oscillator: A Ramsey Interferometry Effect }

\author{A. Bermudez$^1$, M.~A. Martin-Delgado$^1$ and A. Luis$^2$}

\affiliation{ $^1$Departamento de F\'{\i}sica Te\'orica I,
Universidad Complutense, 28040 Madrid, Spain \\$^2$Departamento de
 \'Optica, Universidad Complutense, 28040 Madrid,
Spain}

\begin{abstract}
We study the non-relativistic limit of a paradigmatic model in
Relativistic Quantum Mechanics, the two-dimensional Dirac
oscillator. Remarkably, we find a novel kind of
\emph{Zitterbewegung} which persists in this non-relativistic
regime, and leads to an observable deformation of the particle
orbit. This effect can be interpreted in terms of a Ramsey
Interferometric phenomenon, allowing an insightful connection
between Relativistic Quantum Mechanics and Quantum Optics.
Furthermore, subsequent corrections to the non-relativistic limit,
which account for the usual spin-orbit \emph{Zitterbewegung},
 can be neatly understood in terms of a Mach-Zehnder
 interferometer.
\end{abstract}


\maketitle

\section{Introduction}
\label{sectionI}
 The natural relativistic extension of the quantum
harmonic oscillator, known as the Dirac
oscillator~\cite{moshinsky}, has become a cornerstone in
Relativistic Quantum Mechanics. It was initially introduced as a
relativistic effective model to describe mesons, since it presents
interesting quark-confinement properties~\cite{ito,cook}.
Moreover, subsequent studies have revealed several amazing
properties of the Dirac oscillator in different contexts. Beyond
its exact solvability, the energy spectrum presents  certain
peculiar degeneracies which can be related to a non-trivial
symmetry Lie algebra~\cite{moshinsky_lie}. Furthermore, its
solvability can be traced back to an exact Foldy-Wouthuysen
transformation~\cite{moreno}, and its special properties are
related to a hidden supersymmetry \cite{benitez_susy}.
Additionally, the positive- and negative-energy solutions are
associated to supersymmetrical partners, which ensures the
stability of the Dirac sea under the Dirac oscillator
coupling~\cite{matinez_romero_dirac_sea}.

Some analogies between the dynamics of this relativistic model and
the typical  Jaynes-Cummings (JC) dynamics in Quantum
Optics~\cite{jaynes_cummings} have been discussed
in~\cite{rozmej_1}, and some of its non-relativistic properties
in~\cite{rozmej_2,rozmej_3,rozmej_4}. Remarkably, in a
two-dimensional setup this analogy becomes an exact equivalence
 between the 2+1 Dirac
oscillator Hamiltonian and the Anti-Jaynes-Cummings (AJC)  interaction~\cite{wineland_review}, which
 builds a bridge between two unrelated
fields, Quantum Optics and Relativistic Quantum Mechanics, and
favors a fruitful exchange of ideas between both communities~\cite{bermudez_do_2D}.
Relativistic effects such as the \emph{Zitterbewegung}, a
helicoidal motion performed by the average position of a free
relativistic fermion, can be reinterpreted with the language of
Quantum Optics. This dynamical phenomenon also becomes observable
in the spin and orbital degrees of freedom, where it can be
surprisingly interpreted in terms of optical Rabi oscillations. In
this work we shall be concerned with the extension of this novel
perspective onto the non-relativistic limit of the two-dimensional
Dirac oscillator. In this manner, we are able to identify a novel
feature of the \emph{Zitterbewegung}, which is interpreted as a
Ramsey interferometry effect~\cite{Ramsey}. This remarkable effect
shows that the standard non-relativistic limit described in
Relativistic Quantum Mechanics textbooks should be reconsidered in
this unusual scenario~\cite{greiner}.

This paper is organized as follows: in Sect.~\ref{sectionII}, we
review the properties of the two-dimensional Dirac oscillator and
the exact mapping onto an AJC Hamiltonian. In
Sect.~\ref{sectionIII} the non-relativistic limit is considered
from a Quantum Optics perspective, which allows the prediction of
a novel kind of \emph{Zitterbewegung} which is interpreted as a
Ramsey interferometric phenomenon in Sect.~\ref{sectionIV}. In
this section we also discuss how the effects of this interference
process have strong consequences in the electron trajectory. In
Sect.~\ref{sectionV}, the analysis of additional corrections to
the non-relativistic limit is discussed in terms of a Mach-Zehnder
interferometer. There, we find that the first order non-trivial
correction already shows a perturbative spin-orbit
\emph{Zitterbewegung}~\cite{bermudez_do_2D}. Finally, we conclude
reviewing the consequences of our work in Sect.\ref{sectionVI}.
In appendices~\ref{appendix} and ~\ref{appendixB} we give detailed derivations
of the standard non-relativistic limit
and its complete perturbative series, respectively.

\section{Two-dimensional Dirac Oscillator}
\label{sectionII}

The physical laws that describe the properties of microscopic
particles are accurately described by Quantum Mechanics, and in
particular by the Schr\"{o}dinger equation. Nonetheless, quantum
phenomena occurring at high energies cannot be properly addressed
by such theory, and one must employ Relativistic Quantum
Mechanics~\cite{greiner}. A relativistic spin-1/2 particle of mass
$m$ is described by the Dirac equation
\begin{equation}
\label{ec_dirac}
 \ii\hbar\frac{\partial |\Psi\rangle}{\partial
t}=\left[c\boldsymbol{\alpha}\cdot\textbf{p}+\beta
mc^2\right]|\Psi\rangle,
\end{equation}
where $|\Psi\rangle$ is the four-component Dirac spinor,
$\alpha_i:=\text{off-diag}(\sigma_i,\sigma_i)$ , and
$\beta:=\text{diag}(\mathbb{I}_2,-\mathbb{I}_2)$ are known as the
Dirac matrices which can be expressed in terms of the usual Pauli
matrices $\sigma_i$, $\textbf{p}$ is the momentum operator, and
$c$ stands for the speed of light.

The Dirac oscillator is obtained after the introduction of a
peculiar coupling in the above equation~\eqref{ec_dirac}
\begin{equation}
\label{ec_dirac_oscillator}
 \ii\hbar\frac{\partial |\Psi\rangle}{\partial
t}=\left[c\boldsymbol{\alpha}\cdot\left(\textbf{p}-\ii
m\beta\omega \textbf{r}\right)+\beta mc^2\right]|\Psi\rangle,
\end{equation}
where $\omega$ stands for the Dirac oscillator frequency, and
$\textbf{r}$ represents the particle position. The relativistic
coupling $\textbf{p}-\ii m\beta\omega \textbf{r}$, known as a
Dirac string, cannot be understood as a simple minimal coupling
procedure, and is responsible for the special properties of this
relativistic system. In particular, the non-relativistic limit of
the aforementioned Dirac oscillator~\eqref{ec_dirac_oscillator}
leads to the usual non-relativistic harmonic oscillator with an
additional spin-orbit coupling which shows the intrinsic spin
structure of the relativistic theory~\cite{moshinsky}.

The restriction to two-spatial dimensions appreciably simplifies
the relativistic problem, since the Dirac matrices become $2\times
2$ matrices which can be identified with the so-called Pauli
matrices $\alpha_x=\sigma_x,\alpha_y=\sigma_y,\beta=\sigma_z$. In
this manner, $|\Psi\rangle$ can be described by a 2-component
spinor, and the Dirac oscillator model now takes the form
\begin{equation}
\label{ec_dirac_oscillator_2D}
 \ii \hbar\frac{\partial |\Psi\rangle}{\partial t}=\left[\sum_{j=1}^2c\sigma_{j}\left(p^j-\ii m\sigma_z\omega
r^j\right)+\sigma_z mc^2\right]|\Psi\rangle.
\end{equation}
This two-dimensional system was algebraically solved
\cite{bermudez_do_2D} by introducing chiral
creation and annihilation operators
\begin{equation}
\label{circular_operators}
\begin{array}{c}
  a_r:=\frac{1}{\sqrt{2}}(a_x - \ii a_y),\hspace{2ex}a_r^\dagger:=\frac{1}{\sqrt{2}}(a_x^\dagger + \ii a_y^\dagger) , \\
  a_l:=\frac{1}{\sqrt{2}}(a_x + \ii a_y), \hspace{2ex}a_l^\dagger:=\frac{1}{\sqrt{2}}(a_x^\dagger - \ii a_y^\dagger) , \\
\end{array}
\end{equation}
where $a_x, a_x^\dagger, a_y, a_y^\dagger$, are the usual
annihilation-creation operators
$a^{\dagger}_j=\frac{1}{\sqrt{2}}\left(\frac{1}{\Delta}r^j - \ii
\frac{\Delta}{\hbar}p^j\right)$, and $\Delta=\sqrt{\hbar/m\omega}$
represents the ground state oscillator width. These operators
allow an insightful derivation of the energy spectrum
\begin{equation}
\label{energies}
 E=\pm E_{n_l}=\pm mc^2\sqrt{1+4\xi n_l},
\end{equation}
where the integer $n_l$ stands for the number of left-handed
orbital quanta, and $\xi:=\hbar\omega/mc^2$ is an important
parameter that specifies the importance of relativistic effects in
the Dirac oscillator. A different approach, based on the solution of differential equations, has
also been discussed in~\cite{villalba} .The associated eigenstates are found to be
\begin{equation}
\label{eigenstates}
\begin{array}{c}
  | +E_{n_l} \rangle = \alpha_{n_l} |n_l\rangle \ket{\chi_\uparrow} - \ii \beta_{n_l}|n_l-1\rangle \ket{\chi_\downarrow}, \\
  | -E_{n_l} \rangle=\beta_{n_l}|n_l\rangle \ket{\chi_\uparrow} + \ii \alpha_{n_l}|n_l-1\rangle \ket{\chi_\downarrow}, \\
\end{array}
\end{equation}
where $\ket{\chi_{\uparrow}}:=(1,0)^{t}$ and
$\ket{\chi_{\downarrow}}:=(0,1)^{t}$ are known as the Pauli
spinors,  and $\alpha_{n_l}:=\sqrt{(E_{n_l}+mc^2)/2E_{n_l}}$ and
$\beta_{n_l}:=\sqrt{(E_{n_l}-mc^2)/2E_{n_l}}$ are real
normalization constants.

The notation in Eq.~\eqref{eigenstates} clearly shows that the
relativistic eigenstates exhibit entanglement between the orbital
and spin degrees of freedom. This entanglement property is
essential in order to obtain the relativistic effect of spin-orbit
\emph{Zitterbewegung}, where certain oscillations in the orbital
and spin angular momentum are unambiguously identified with the
interference of positive- and negative-energy components.
Introducing the spin and angular momentum operators $S_z=\half
\hbar\sigma_z$ and $L_z=\hbar(a^{\dagger}_ra_r-a^{\dagger}_la_l)$,
and considering the initial state $|\Psi(0)\rangle:=|n_l-1\rangle
\ket{\chi_\downarrow}$, one immediately finds the so-called
spin-orbit oscillations associated to the \emph{Zitterbewegung}
\begin{equation}
\label{zitter_dynamics}
\begin{array}{l}
  \langle L_z\rangle_{t}=-\frac{4\xi n_l}{1+4\xi
n_l}\hbar\sin^2\omega_{n_l}t -\hbar(n_l-1), \\
    \langle S_z\rangle_t=\hspace{2ex}\frac{4\xi n_l}{1+4\xi
n_l}\hbar\sin^2\omega_{n_l}t -\frac{\hbar}{2},  \\
\langle J_z\rangle_t=\hspace{2ex}\hbar(\half - n_l),
\end{array}
\end{equation}
where $J_z=L_z+S_z$ stands for the $z$-component of the total
angular momentum and is obviously a conserved quantity.

Finally, we recall the interesting mapping between the
two-dimensional Dirac oscillator and the AJC model
\cite{bermudez_do_2D}, where the relativistic hamiltonian can
be written as
\begin{equation} \label{AJC_DO}
\begin{split}
H&=\hbar(g\sigma^+a_l^\dagger+g^*\sigma^-a_l)+\delta\sigma_z,
\end{split}
\end{equation}
with $\sigma^+:=\ket{\chi_{\uparrow}}\bra{\chi_{\downarrow}}$,
$\sigma^-:=\ket{\chi_{\downarrow}}\bra{\chi_{\uparrow}}$ as the
spin raising and lowering operators, $g := 2 \ii mc^2 \sqrt{\xi} /
\hbar$ as the coupling strength between orbital and spin degrees
of freedom, and where $\delta:=mc^2$ can be interpreted as a
detuning parameter. In Quantum Optics, this Hamiltonian describes
an Anti-Jaynes-Cummings interaction, and can be implemented with
trapped ions \cite{wineland_review}. Within this novel
perspective, the electron spin can be associated with a two-level
atom, and the orbital circular quanta with the ion quanta of
vibration, i.e., phonons. Note that there is also  the possibility
to map this relativistic Hamiltonian onto the more standard
Jaynes-Cummings model~\cite{comment1}. In the following section,
we derive the non-relativistic limit of the two-dimensional Dirac
oscillator~\eqref{AJC_DO}, and discuss the nature of the physical
properties described in
Eqs.~\eqref{energies}-\eqref{zitter_dynamics} in this
non-relativistic scenario.

\section{Non-relativistic limit in Quantum Optics}
\label{sectionIII}

\begin{figure}[!hbp]

\centering

\begin{overpic}[width=8.5cm]{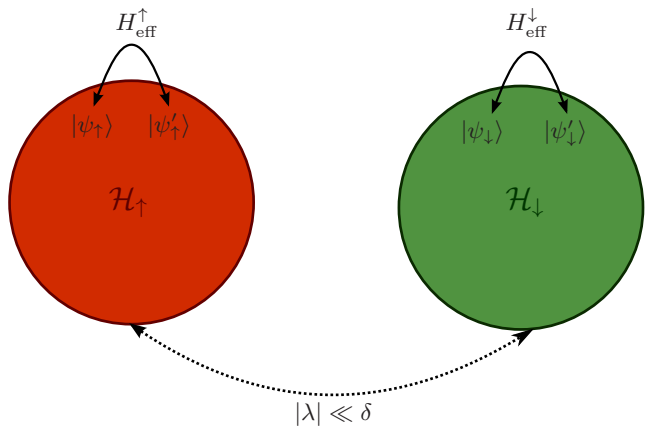}

\put(16,38){\textcolor[rgb]{0.37,0.00,0.00}{\large{$\mathcal{H}_{\uparrow}$}}}
\put(78,38){\textcolor[rgb]{0.00,0.21,0.00}{\large{$\mathcal{H}_{\downarrow}$}}}
\put(10,50){$\ket{\psi_{\uparrow}}$}\put(22,50){$\ket{\psi'_{\uparrow}}$}
\put(71,49){$\ket{\psi_{\downarrow}}$}\put(84,49){$\ket{\psi'_{\downarrow}}$}
\put(45,5){$|\lambda|\ll\delta$}
\put(17,66){$H_{\text{eff}}^{\uparrow}$}
\put(78,66){$H_{\text{eff}}^{\downarrow}$}
\end{overpic}

\caption{"(Color
 online)" Schematic  diagram of the Dirac oscillator couplings in
the non-relativistic limit. The subspaces $\mathcal{H}_{\uparrow}$
and $\mathcal{H}_{\downarrow}$ become disconnected in this
regime}\label{couplings}

\end{figure}

The original Quantum Optics perspective of the two-dimensional
Dirac oscillator in Eq.~\eqref{AJC_DO} stimulates the use of
quantum optical tools in  a relativistic quantum framework, and
viceversa. In particular, we can use the quasi-degenerate
perturbation theory~\cite{cohen} in order to derive an effective
Hamiltonian in the non-relativistic limit. This regime is attained
when the relativistic parameter fulfills $\xi n_l\ll1$, which
allows the usual description of the Dirac oscillator Hamiltonian
in Eq.~\eqref{AJC_DO}
\begin{equation}
H=H_0+\lambda V,
\end{equation}
where $H_0$ is the unperturbed Hamiltonian
\begin{equation}
H_0=\delta\sigma_z,
\end{equation}
and $\lambda V$ represents the following perturbation
\begin{equation}
\label{perturbation} \lambda V=
\lambda(\sigma^+a_l^\dagger-\sigma^-a_l),
\end{equation}
where the interaction coupling $\lambda:=2\ii mc^2\sqrt{\xi}$
satisfies $|\lambda|\ll\delta$, and consequently the coupling in
Eq.~\eqref{perturbation} can be treated as a small perturbation.
In this regime, the Hilbert space can be divided into two
approximately disconnected subspaces
\begin{equation}
\mathcal{H}\approx\mathcal{H}_{\uparrow}\oplus\mathcal{H}_{\downarrow},
\end{equation}
where
\begin{equation}
\label{invariant_subspaces}
\begin{split}
\mathcal{H}_{\uparrow}=\text{span}\{\ket{n_l}\ket{\chi_{\uparrow}}; n_l=0,1...\},\\
\mathcal{H}_{\downarrow}=\text{span}\{\ket{n_l}\ket{\chi_{\downarrow}};n_l=0,1...\}.
\end{split}
\end{equation}
This is easily understood from the fact that $\xi n_l\ll 1$ is
equivalent to $\hbar\omega\ll\delta$, which implies that the
perturbation in Eq.~\eqref{perturbation} does not suffice to
induce transitions between the spinor components. In this manner,
the subspaces corresponding to the spinor degrees of freedom
become decoupled, and we may describe the effective dynamics in
such subspaces ( see fig.~\ref{couplings} ).

 In order to obtain the effective Hamiltonian, we
rewrite the perturbation in Eq.~\eqref{perturbation} as follows
\begin{equation}
\lambda V=\sum_{\mu}\lambda_{\mu}A_{\mu}B_{\mu},
\end{equation}
where $\mu=1,2$, the operators $A_1=a_{l},A_2=a^{\dagger}_l$
describe the slow varying orbital degrees of freedom, whereas the
operators $B_1=\sigma^-,B_2=\sigma^+$ represent the coupling of
the fast spinorial degrees of freedom, and
$\lambda_1=g,\lambda_2=g^*$. The effective Hamiltonians correspond
to
\begin{equation}
\begin{split}
H_{\text{eff}}^{\uparrow}&=\hspace{2ex}\delta+\sum_{\mu\mu'}\bra{\chi_{\uparrow}}\lambda_{\mu}B_{\mu}\frac{1}{\delta-H_0}\lambda_{\mu'}B_{\mu'}\ket{\chi_{\uparrow}}A_{\mu}A_{\mu'},\\
H_{\text{eff}}^{\downarrow}&=-\delta+\sum_{\mu\mu'}\bra{\chi_{\downarrow}}\lambda_{\mu}B_{\mu}\frac{1}{\delta-H_0}\lambda_{\mu'}B_{\mu'}\ket{\chi_{\downarrow}}A_{\mu}A_{\mu'},
\end{split}
\end{equation}
which can be readily evaluated as
\begin{equation}
\label{NR_effective_hamiltonian_QO}
H_{\text{eff}}=\left[%
\begin{array}{cc}
  mc^2+2\hbar\omega a_l^{\dagger}a_l & 0 \\
  0 & -mc^2-2\hbar\omega a_la_l^{\dagger} \\
\end{array}%
\right].
\end{equation}
The effect of the relativistic corrections in this regime can be
understood as a level shift with respect to the rest mass energy
that depends on the number of left-handed quanta. The
non-relativistic energies associated to the corresponding
eigenstates $\ket{n_l}\ket{\chi_{\uparrow}}$ and
$\ket{n_l-1}\ket{\chi_{\downarrow}}$ are
\begin{equation}
\label{energies_NR}
\begin{split}
E_{\ket{n_l}\ket{\chi_{\uparrow}}}&=+mc^2(1+2\xi n_l), \\
E_{\ket{n_l-1}\ket{\chi_{\downarrow}}}&=-mc^2(1+2\xi n_l),
\end{split}
\end{equation}
which are equivalent to the leading order correction of the exact
eigenvalues~\eqref{energies}  in the limit $\xi n_l\ll1$, namely,
\begin{equation}
E=\pm mc^2\sqrt{1+4\xi n_l}\approx\pm mc^2(1+2\xi
n_l)+\mathcal{O}\left((\xi n_l)^2\right).
\end{equation}
Therefore, we obtain the so-called  energy shift term $\Delta
E=2mc^2(1+2\xi n_l)$ , which is usually known as a dynamical Stark
shift term in the quantum optics literature~\cite{Stark}. In
Optics, the non-relativistic effective
Hamiltonian~\eqref{NR_effective_hamiltonian_QO} achieved for large
enough detuning is equivalent to the dispersive linear
susceptibility and the real part of the refraction index, with
opposite contributions from the excited and ground states
\cite{SI86}.

This noteworthy interpretation of the non-relativistic limit in
terms of measurable optical quantities is shown to be equivalent
to the standard non-relativistic limit in Relativistic Quantum
Mechanics in Appendix~\ref{appendix}. In the following section, we
shall use this remarkable perspective to describe a novel sign of
the \emph{Zitterbewegung}, which can be understood in terms of a
Ramsey interferometry effect~\cite{Ramsey}.

\section{\emph{Zitterbewegung} in the non-relativistic limit}
\label{sectionIV}

As discussed in previous sections, the interference between
positive- and negative-energy components gives rise to a
relativistic oscillatory behavior known as \emph{Zitterbewegung}.
This phenomenon has a pure relativistic nature, and therefore it
is usually believed to vanish in the non-relativistic limit.
Contrary to common belief, we show in this section how a peculiar
\emph{Zitterbewegung} effect can still arise  in the
non-relativistic regime if the initial state is appropriately
prepared. Furthermore, we also discuss how this dynamics might be
interpreted as a Ramsey interferometry phenomenon, and how it can
lead to measurable effects in the particle trajectory.

Let us consider an initial state
$\ket{\Psi(0)}:=\alpha\ket{n_l}\ket{\chi_{\uparrow}}+\beta\ket{n_l-1}\ket{\chi_{\downarrow}}$,
which involves both spinorial components, where
$\alpha,\beta\in\mathbb{C}$ are correctly normalized
$|\alpha|^2+|\beta|^2=1$. One directly observes that this state
mixes the positive- and negative-energy solutions in
Eq.~\eqref{energies_NR}, which is the fundamental ingredient
leading to the \emph{Zitterbewegung}. In order to obtain such
effect, we derive the time evolution under the effective
Hamiltonian~\eqref{NR_effective_hamiltonian_QO}
\begin{equation}
\ket{\Psi(t)}=\alpha
e^{-\ii\Omega_{n_l}t}\ket{n_l}\ket{\chi_{\uparrow}}+\beta
e^{+\ii\Omega_{n_l}t}\ket{n_l-1}\ket{\chi_{\downarrow}},
\end{equation}
where $\Omega_{n_l}:=mc^2(1+2\xi n_l)/\hbar$. The corresponding
spin-orbit expected values~\eqref{zitter_dynamics} become
\begin{equation}
\begin{array}{l}
  \langle L_z\rangle_{t}=-\hbar\left(n_l+|\beta|^2\right), \\
    \langle S_z\rangle_t=\frac{\hbar}{2}\left(|\alpha|^2-|\beta|^2\right),  \\
\langle
J_z\rangle_t=\frac{\hbar}{2}\left(2n_l+|\alpha|^2-3|\beta|^2\right),
\end{array}
\end{equation}
where any remainder of the original oscillatory
\emph{Zitterbewegung} in Eq.~\eqref{zitter_dynamics} has
completely vanished. Nevertheless, positive- and negative-energy
components are simultaneously involved in the initial state, and
therefore there must exist some kind of novel
\emph{Zitterbewegung}.

The possibility to observe such effect can be achieved if we
consider the superposition of states with different energy modulus
$|E_{n_l}|$. This novel \emph{Zitterbewegung} can be exemplified
by a Ramsey interferometer in which the field is prepared in a
coherent state and the atom in a 50 \% superposition of its
excited and ground states ( i.e.
$\alpha\ket{\chi_{\uparrow}}+\beta\ket{\chi_{\downarrow}}$) by
resonant interaction with a classical laser beam~\cite{Ramsey}.
Afterwards, the effective interaction produces a different field
evolution conditioned to the atomic level. Finally, resonant
interaction of the atom with a second laser beam mixes the
contributions from the upper and lower levels leading to the
interference of the positive- and negative-energy evolutions.
Therefore, the Ramsey fringes such as the ones in Refs.~\cite{RF}
can be regarded as suitable examples of \emph{Zitterbewegung} in
this particular regime.

 Let us exemplify this Ramsey-\emph{Zitterbewegung} with the
 following initial state prepared in a superposition of the
 two-spinor states
$\ket{\Psi(0)}:=\ket{z_l}(\alpha\ket{\chi_{\uparrow}}+\beta\ket{\chi_{\downarrow}})$,
where
$\ket{z_l}:=\ee^{-|z_l|^2/2}\sum_{n_l}z_l^{n_l}\ket{n_l}/\sqrt{n_l!}$
 represents a left-handed coherent state with $z_l\in\mathbb{C}$.
This initial state also involves  positive- and negative-energy
components, and its time evolution is
\begin{equation}
\label{coherent_evolution}
 \ket{\Psi(t)}=\alpha
\ee^{-\ii\Omega_0t}\ket{z_l\ee^{-2\ii\omega
t}}\ket{\chi_{\uparrow}}+\beta
\ee^{+\ii\Omega_1t}\ket{z_l\ee^{+2\ii\omega
t}}\ket{\chi_{\downarrow}}.
\end{equation}
As time elapses, the phase evolution of the  orbital coherent
state is strongly correlated to the internal spinorial degree of
freedom, just as the Ramsey interferometric time evolution. This
peculiar correlated dynamics is a clear consequence of the
coexistence of positive- and negative-energy modes in the initial
state, and therefore it stands as a direct symptom of
\emph{Zitterbewegung}. The final step in the Ramsey
interferometric experiment is to recombine both spinor components
leading to the interference of positive- and negative-energy
modes. The measurement of $S_z$ after this mixing effect is
equivalent to the measurement of the $x$-component of the spin
angular momentum $S_x:=\frac{\hbar}{2}\sigma_x$ in the
state~\eqref{coherent_evolution}
\begin{equation}
\langle
S_x\rangle_t=\hbar\mathcal{V}(t)\cos\left[(\Omega_0+\Omega_1)t+|z_l|^2\sin(4\omega
t)+\text{arg}({\alpha^*\beta})\right]
\end{equation}
where
$\mathcal{V}(t)=|\alpha^*\beta|\ee^{-2|z_l|^2\sin(2\omega
t)}$ can be identified with a periodic visibility factor which
precedes the desired Ramsey interference term. Therefore, an
oscillatory behaviour in the $x-$spin component can be directly
traced back to a \emph{Zitterbewegung} in the non-relativistic
limit. Note that this phenomenon is consequence of an appropriate
preparation of the initial state, involving both energy modes,
rather than the pure dynamical effect in
Eqs.~\eqref{zitter_dynamics}.

\begin{figure}[!hbp]

\centering

\begin{overpic}[width=8.5cm]{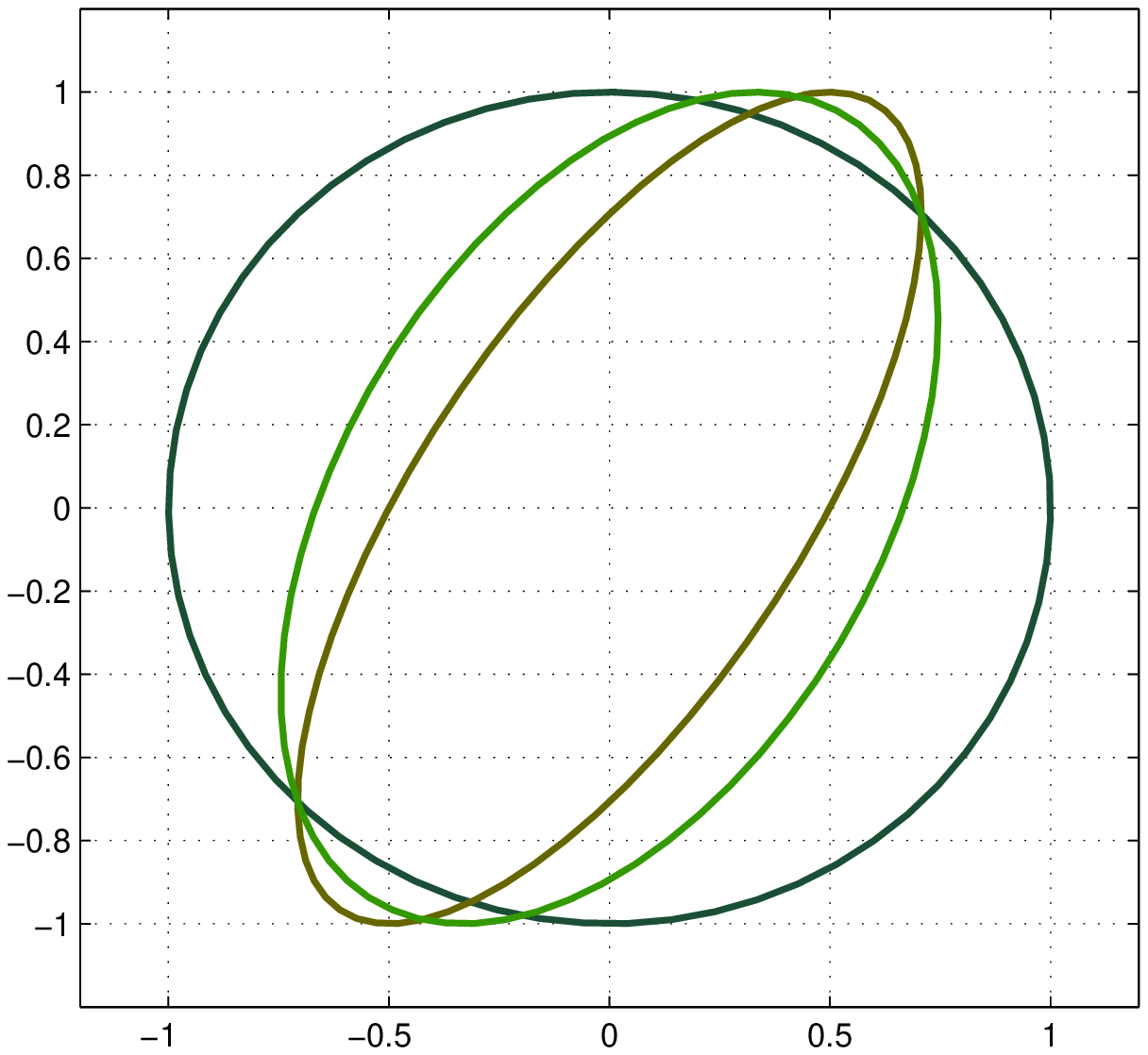}
\put(45,3){$\langle x \rangle /|z_l|^2$} \put(-5,45){$\langle
y\rangle /|z_l|^2$}
\put(73,21){\textcolor[rgb]{0.00,0.21,0.00}{$\small{\alpha_1}$}}
\put(65,28){\textcolor[rgb]{0.41,0.70,0.13}{$\small{\alpha_2}$}}
\put(51,38){\textcolor[rgb]{0.50,0.50,0.00}{$\small{\alpha_3}$}}
\end{overpic}

\caption{"(Color
 online)" Electron trajectory for different initial states where
the coherent state phase is $\phi_l=\pi/2$, and
$\alpha_{i}=\{1,\sqrt{2/3},\sqrt{1/2}\}$, with $i=1,2,3$. Note
that $\alpha_1=1$ corresponds to the circular
\emph{Zitterbewegung}-free evolution in
Eq.~\eqref{ZB_free_elecron_dynamics}, as compared to the elliptic
orbits caused by a \emph{Zitterbewegung} interference.
}\label{electron_trajectory}

\end{figure}

This Ramsey interferometric dynamics~\eqref{coherent_evolution}
can lead to measurable effects in the electron trajectory, since
the interference of both energy modes undeniably causes a
deformation of the particle orbit. The electron trajectory is
described by the following position operators $
x=\Delta(a_r+a_r^{\dagger}+a_l+a_l^{\dagger})/2$,
$y=\ii\Delta(a_r-a_r^{\dagger}-a_l+a_l^{\dagger})/2$, whose
expectation values evolve according to
\begin{equation}
\label{ZB_elecron_dynamics}
\begin{split}
\langle x\rangle_t&=\Delta|z_l|\left[|\alpha|^2\cos(2\omega
t-\phi_l)+|\beta|^2\cos(2\omega t+\phi_l)\right],\\
\langle
y\rangle_t&=\Delta|z_l|\left[|\beta|^2\hspace{0.2ex}\sin(2\omega
t+\phi_l)-|\alpha|^2\hspace{0.2ex}\sin(2\omega t-\phi_l)\right],
\end{split}
\end{equation}
where we have used $z_l=|z_l|\ee^{\ii\phi_l}$. The particle
trajectory described in Eq.~\eqref{ZB_elecron_dynamics} has a
remarkable periodic character, and must be compared to the
\emph{Zitterbewegung-free} trajectory of an initial state
$\ket{\Psi(0)}:=\ket{z_l}\ket{\chi_{\uparrow}}$, which is
described by a circular orbit
\begin{equation}
\label{ZB_free_elecron_dynamics}
\begin{split}
\langle x\rangle_t&=+\Delta|z_l|\cos(2\omega
t-\phi_l),\\
\langle y\rangle_t&=-\Delta|z_l|\hspace{0.2ex}\sin(2\omega
t-\phi_l).
\end{split}
\end{equation}
Comparing both trajectories in
Eqs.~\eqref{ZB_elecron_dynamics},\eqref{ZB_free_elecron_dynamics},
we realize that the \emph{Zitterbewegung} phenomenon leads to a
deformation of the electron circular
orbit~\eqref{ZB_free_elecron_dynamics} ( see
fig.~\ref{electron_trajectory})

In light of the results presented in this section, we may conclude
that \emph{Zitterbewegung} phenomena may also arise in a
non-relativistic regime as long as the initial state involves both
energy modes, which is still a relativistic property. The initial state, which can be described by
a coherent superposition of positive- and negative-energy solutions, cannot be described
in the realm of non-relativistic quantum mechanics. Therefore,
the persistence of \emph{Zitterbewegung} in the non-relativistic regime
 can be traced back to the relativistic nature of the initial state,
 when carefully prepared.

Furthermore, we have also described how this interference effect
can be unexpectedly interpreted in terms of Ramsey fringes in the
context of Quantum Optics. The effect of this relativistic
interference can be observed in oscillations in the $x-$spin
component, or even more drastically in the deformation of the
electron circular orbit into an elliptic trajectory.

\section{Corrections to the non-relativistic limit}
\label{sectionV}

In section~\ref{sectionIII}  we have discussed the
non-relativistic limit of the two-dimensional Dirac oscillator,
where the decoupling of the spinor components leads to a dynamical
Stark shift in the energy levels. It was precisely this energy
shift, which allowed the surprising description of the
\emph{Zitterbewegung} in terms of a Ramsey interferometry effect
in Sect.~\ref{sectionIV}. In the present section, we consider how
relativistic effects modify such picture as the parameter $\xi$
increases, and interpret the usual spin-orbit
\emph{Zitterbewegung} described in Eqs.~\eqref{zitter_dynamics} as
a first order perturbation term. The picture developed in
section~\ref{sectionIII} is no longer valid, since it assumes the
decoupling of the spinor subspaces~\eqref{invariant_subspaces},
which forbids this peculiar dynamics. Therefore, we discuss a
novel description which allows the interpretation of the
spin-orbit \emph{Zitterbewegung} in terms of a Mach-Zehnder
interferometer.

\begin{figure}[!hbp]

\centering

\begin{overpic}[width=8cm]{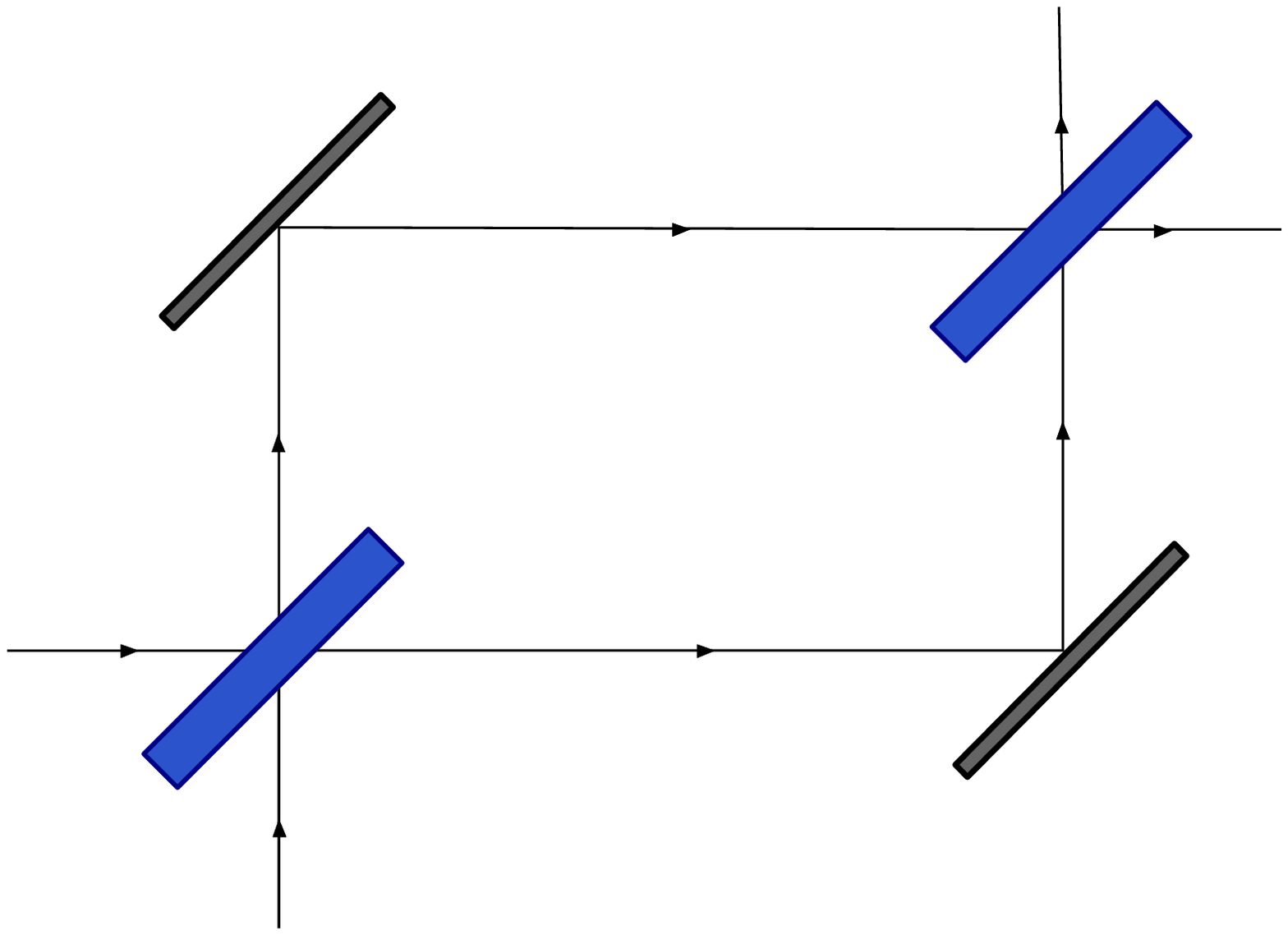}
\put(24,30){\textcolor[rgb]{0.00,0.00,0.35}{$\theta_{n_l}$}}
\put(70,71){\textcolor[rgb]{0.00,0.00,0.35}{$-\theta_{n_l}$}}
\put(3,38){$\ket{n_l}\ket{\chi_{\uparrow}}$}
\put(84,82){$\ket{n_l}\ket{\chi_{\uparrow}}$}
\put(23,50){$\ket{+E_{n_l}}$} \put(66,50){$\ket{-E_{n_l}}$}
 \put(45,62){$\ket{+E_{n_l}}$}
\put(23,18){$\ket{n_l-1}\ket{\chi_{\downarrow}}$}
\put(45,38){$\ket{-E_{n_l}}$}
\put(84,62){$\ket{n_l-1}\ket{\chi_{\downarrow}}$}
\end{overpic}

\caption{"(Color
 online)" Mach-Zehnder interferometer diagram of the Dirac
oscillator evolution operator}\label{mach-zender_bis}

\end{figure}

Let us consider the AJC-mapping of the two-dimensional Dirac
oscillator~\eqref{AJC_DO}, which allows the description of the
Hilbert space as a series of invariant subspaces
$\mathcal{H}=\bigoplus_{n_l=1}^{\infty}\mathcal{H}_{n_l}$, where
\begin{equation}
\mathcal{H}_{n_l}=\text{span}\{\ket{n_l}\ket{\chi_{\uparrow}},\ket{n_l-1}\ket{\chi_{\downarrow}}\}.
\end{equation}
The relativistic Hamiltonian in these subspaces reads
\begin{equation}
 H_{n_l}= mc^2 \left (  \sigma_z - \eta_{n_l} \sigma_y \right ) ,
\end{equation}
where we have introduced  a parameter $ \eta_{n_l} = 2 \sqrt{\xi
n_l}$  directly related to the small relativistic parameter $\xi
n_l\ll 1$. This interaction can be considered as a rotation
of the $\sigma_z$ term along the $x$-axis
\begin{equation}
 H_{n_l} =  mc^2 \sqrt{1 + \eta_{n_l}^2} \ee^{  -\ii
 \theta_{n_l}
\sigma_x } \sigma_z \ee^{  \ii \theta_{n_l} \sigma_x },
\end{equation}
where the rotation angle satisfies $\tan (2 \theta_{n_l}) :=
\eta_{n_l} $~\cite{KS00}. The unitary  time evolution operator can
be expressed as follows
\begin{equation}
\label{evolution_operator} U_{n_l} = \ee^{  -\frac{\ii t}{\hbar} H_{n_l}
} = \ee^{ - \ii \theta_{n_l} \sigma_x} \ee^{-\ii
\phi_{n_l}(t)  \sigma_z } \ee^{\ii \theta_{n_l} \sigma_x },
\end{equation}
where
\begin{equation}
\label{dephasing}
 \phi_{n_l}(t) = \frac{mc^2}{\hbar} \sqrt{1 + \eta_{n_l}^2}t.
 \end{equation}
This evolution has a clear interferometric interpretation in terms
of a Mach-Zehnder interferometer ( see
fig.~\ref{mach-zender_bis}).

\begin{figure}[!hbp]

\centering

\begin{overpic}[width=8cm]{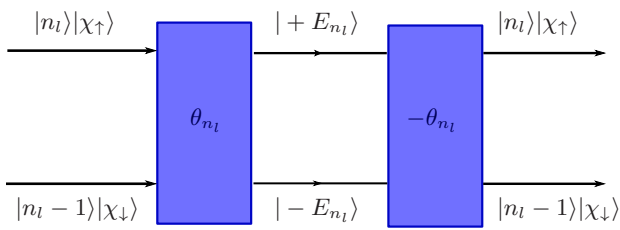}
\put(31,30){\textcolor[rgb]{0.00,0.00,0.35}{$\theta_{n_l}$}}
\put(67,30){\textcolor[rgb]{0.00,0.00,0.35}{$-\theta_{n_l}$}}
\put(5,46){$\ket{n_l}\ket{\chi_{\uparrow}}$}
\put(82,46){$\ket{n_l}\ket{\chi_{\uparrow}}$}
\put(45,46){$\ket{+E_{n_l}}$}
\put(2,15){$\ket{n_l-1}\ket{\chi_{\downarrow}}$}
\put(45,15){$\ket{-E_{n_l}}$}
\put(82,15){$\ket{n_l-1}\ket{\chi_{\downarrow}}$}
\end{overpic}

\caption{"(Color
 online)" Schematic Mach-Zehnder interferometer diagram of the
Dirac oscillator evolution operator}\label{mach-zender}

\end{figure}

We can understand the interferometric process clearer in the
three-step process of fig.~\ref{mach-zender}.  Here, the term
$e^{\ii \theta_{n_l} \sigma_x}$ represents a beam splitter at the
entrance of the interferometer
\begin{equation}
\ee^{\ii \theta_{n_l} \sigma_x}=\left[%
\begin{array}{cc}
  \cos \theta_{n_l} & - \ii \sin \theta_{n_l} \\
\cr - \ii \sin \theta_{n_l} &  \cos\theta_{n_l} \\
\end{array}%
\right],
\end{equation}
the following term $\ee^{-\ii \phi_{n_l}(t)  \sigma_z }$ describes
the dephasing process in the two arms of the interferometer
\begin{equation}
\ee^{-\ii \phi_{n_l}(t)  \sigma_z }=\left[%
\begin{array}{cc}
  \ee^{-\ii \phi_{n_l}}  & 0 \\
0 &  \ee^{+\ii \phi_{n_l}} \\
\end{array}%
\right],
\end{equation}
 and the remaining term $\ee^{ - \ii \theta_{n_l} \sigma_x}$ stands
 for the final beam splitter which produces the interference
 between the dephased beams that have traveled through different
 paths of the interferometer
\begin{equation}
\ee^{ - \ii \theta_{n_l} \sigma_x}=\left[%
\begin{array}{cc}
  \cos \theta_{n_l} &  \ii \sin \theta_{n_l} \\
 \ii \sin \theta_{n_l} &  \cos\theta_{n_l} \\
\end{array}%
\right].
\end{equation}
Remarkably enough, this three-step process captures the essence of
the relativistic dynamical properties in the two-dimensional Dirac
oscillator. The two incoming beams might be interpreted as the
upper and lower components of the relativistic spinor. The first
beam splitter is responsible for the mixture of these components
so that the two arms of the interferometer can be associated to
positive- and negative-energy solutions~\eqref{eigenstates}. In
such manner, the time evolution inside the interferometer can be
understood as a phase shift between the positive- and
negative-energy solutions since their phases evolve with opposite
sign. Finally, the second beam splitter is responsible for the
interference of the interferometer beams, which consequently
represents the interference of positive- and negative-energy
solutions. This is exactly the essence of the
\emph{Zitterbewegung} in Relativistic Quantum
Mechanics~\cite{bermudez_do_2D}, which can be surprisingly
identified with a simple interferometric mechanism for the
two-dimensional Dirac oscillator.

\vspace{2ex}

In the non-relativistic regime described in
section~\ref{sectionIII}, the interferometric picture is
significantly simplified. In this limit, the parameter $\eta_{n_l}$ satisfies $\eta_{n_l}
\ll 1$  for significative values of the initial number of orbital quanta
$n_l$ . Therefore, we can approximate
\begin{equation}
\sqrt{1+\eta_{n_l}^2} \approx 1 + \frac{1}{2} \eta_{n_l}^2 ,
\hspace{2ex} \ee^{ \ii \theta_{n_l} \sigma_x} \approx \left[%
\begin{array}{cc}
  1 & 0 \\
  0 & 1 \\
\end{array}%
\right],
\end{equation}
and thus the rotation angle $\theta_{n_l}\approx0$ becomes
vanishingly small. In this manner, the action of the two beam
splitters in fig.~\ref{mach-zender_bis} is negligible, and the
non-trivial remaining  effect is the dephasing of the upper and
lower components of the relativistic spinor in
Eq.~\eqref{dephasing}, which yields
\begin{equation}
\label{frequencies}
 \phi_{n_l}(t) \approx
\left[\frac{mc^2}{\hbar} + 2 \omega n_l\right] t=:\Omega_{n_l}t.
\end{equation}
 As discussed
in the preceding section, these phase shifts can manifest
themselves in a Ramsey interferometer providing a practical
realization of \emph{Zitterbewegung}. Beyond this example that
requires the preparation of the atom in a superposition state, we
can obtain a further example of \emph{Zitterbewegung} dynamically
induced by a first-order relativistic correction.

The remarkable advantage of the interferometric interpretation
developed in this section is the possibility to go beyond this
non-relativistic regime, and consider higher-order relativistic
corrections to the aforementioned dynamics. In order to do so, we
retain the relativistic corrections up to the following order
 \begin{equation}
\sqrt{1+\eta_{n_l}^2} \approx 1 + \frac{1}{2} \eta_{n_l}^2 ,
\hspace{2ex}
e^{\ii \theta_{n_l} \sigma_x} \approx \left[%
\begin{array}{cc}
  1 - \frac{\eta^2_{n_l} }{8} & \ii \frac{\eta_{n_l} }{2} \\
\ii \frac{\eta_{n_l} }{2} & 1 - \frac{\eta^2_{n_l} }{8} \\
\end{array}%
\right] ,
\end{equation}
which leads to the following evolution operator
\begin{equation}
U_{n_l} \approx \left[%
\begin{array}{cc}
  \ee^{ -\ii \phi_{n_l}} - \frac{\ii \eta^2_{n_l} }{2}\sin \phi_{n_l} & - \eta_{n_l} \sin \phi_{n_l} \\
 \eta_{n_l} \sin \phi_{n_l} &  \ee^{ +\ii \phi_{n_l}} + \frac{\ii \eta^2_{n_l} }{2} \sin \phi_{n_l} \\
\end{array}%
\right] .
\end{equation}
Note that the rotation angle is no longer negligible, and the
action of the beam splitters becomes noticeable up to the
perturbative order considered so far. The interference between the
positive- and negative-energy solutions appears as a pure
relativistic effect, leading to the spin-orbit
\emph{Zitterbewegung} if we consider an initial state
$|\Psi(0)\rangle:=|n_l-1\rangle \ket{\chi_\downarrow}$
\begin{equation}
\label{zitter_dynamics_NR}
\begin{array}{l}
  \langle L_z\rangle_t=-4\xi n_l\hbar\sin^2\Omega_{n_l}t-\hbar(n_l-1)+\mathcal{O}\left((\xi n_l)^2\right), \\
  \langle S_z\rangle_t=\hspace{2ex}4\xi n_l\hbar\sin^2\Omega_{n_l}t -\frac{\hbar}{2}+\mathcal{O}\left((\xi n_l)^2\right), \\
  \langle J_z\rangle_t=\hspace{2ex}\hbar(\half - n_l),
\end{array}
\end{equation}
which clearly coincide with the expansion on the small parameter
$\xi n_l\ll1$ of the dynamical evolution described in
Eqs.~\eqref{zitter_dynamics}. The oscillations in the angular
momentum observables are therefore a direct consequence of the
interference between positive- and negative-energy solutions
introduced by the Mach-Zehnder beam splitters. The visibility of
the interference phenomenon in the spin degrees of freedom is
\begin{equation}
\mathcal{V}=\left|\frac{\langle S_z\rangle_{\text{max}}-\langle
S_z\rangle_{\text{min}}}{\langle S_z\rangle_{\text{max}}+\langle
S_z\rangle_{\text{min}}}\right|\approx4\xi
n_l+\mathcal{O}\left((\xi n_l)^2\right),
\end{equation}
which clearly fulfils $\mathcal{V}\ll1$ at this level of
perturbation theory. As we discuss in appendix~\ref{appendixB},
the visibility of these \emph{Zitterbewegung} oscillations increases considerably as
relativistic effects become more pronounced, and subsequent
perturbative orders are taken into account.

\begin{figure}[!hbp]

\centering

\begin{overpic}[width=7.5cm]{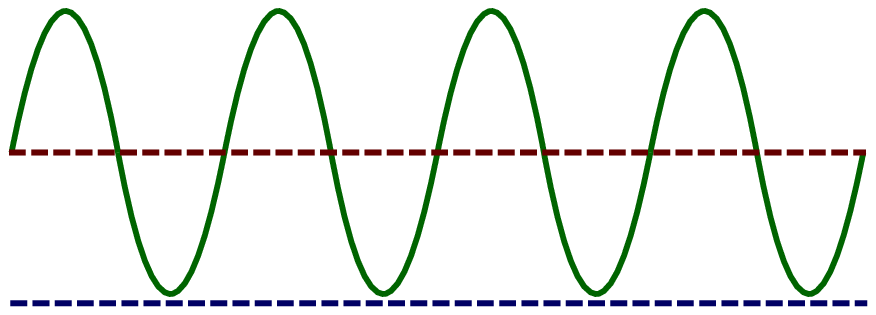}
\put(86,45){\textcolor[rgb]{0.00,0.29,0.15}{$\langle
L_z\rangle_t$}}
\put(93,35){\textcolor[rgb]{0.50,0.25,0.25}{$\overline{\langle
L_z\rangle_t}$}}
\put(83,8){\textcolor[rgb]{0.00,0.00,0.47}{$-\hbar(n_l-1)$}}

\end{overpic}

\caption{"(Color
 online)" Effective frequency displacement of a non-relativistic
harmonic oscillator due to first order relativistic
corrections}\label{desplazamiento_frecuencia}

\end{figure}

 Additionally, one must
also consider the difference in the order of magnitude of the
superposed frequencies in Eq.~\eqref{frequencies}, where $2\omega
n_l\ll mc^2/\hbar$, which makes it difficult to observe the
aforementioned instantaneous oscillations. In such case, we can
also perform a time average of Eqs.~\eqref{zitter_dynamics_NR}
\begin{equation}
\label{timeavg_zitter_dynamics_NR}
\begin{array}{l}
  \overline{\langle L_z\rangle_t}=-2\xi n_l\hbar-\hbar(n_l-1)+\mathcal{O}\left((\xi n_l)^2\right), \\
  \overline{\langle S_z\rangle_t}=\hspace{2ex}2\xi n_l\hbar -\frac{\hbar}{2}+\mathcal{O}\left((\xi n_l)^2\right), \\
  \overline{\langle J_z\rangle_t}=\hspace{2ex}\hbar(\half - n_l),
\end{array}
\end{equation}
which can be readily interpreted as a frequency shift in a
non-relativistic left-handed harmonic oscillator
$\omega\to\omega+2\xi$ ( see
fig.~\ref{desplazamiento_frecuencia}).

\section{Conclusions}
\label{sectionVI}

In this paper we have considered the intriguing relativistic
\emph{Zitterbewegung} in the two-dimensional Dirac oscillator from
an interferometric point of view. The exact mapping between the
relativistic model and the Anti-Jaynes-Cummings interaction,
suggests the use of quantum optical tools in the study of
relativistic quantum phenomena. In this sense, the
non-relativistic limit of the Dirac oscillator can be understood
as a Ramsey interferometric effect, and  interesting
\emph{Zitterbewegung}-dynamics arise when the initial state is
carefully prepared. Actually, we have described how
\emph{Zitterbewegung-free} circular orbits become elliptic
trajectories due to the interference of positive- and
negative-energy modes. This insightful interferometric
interpretation can be carried further on to subsequent
relativistic corrections, in terms of a Mach-Zehnder
interferometric process as shown in appendix~\ref{appendixB}. The
effect of the Mach-Zehnder beam splitters is responsible of the
spin-orbit \emph{Zitterbewegung}, and becomes more relevant as the
relativistic parameter is increased.

It is interesting to point out that all these exotic relativistic
effects may be observed in an ion-trap tabletop experiment
following the proposal described in~\cite{bermudez_do_2D}. In this
experimental setting, the relativistic parameter can attain all
possible values regarding current technology possibilities. This
fact should allow the experimentalist to study this
non-relativistic regime and the interferometric effects discussed
in this work.

Finally, we would also like to stress that an exciting dialogue
between Quantum Optics and Relativistic Quantum Mechanics
scientists can be performed in the light of our results.
These two communities can collaborate
in order to offer a different perspective to archetypical
phenomena of both disciplines.

\noindent{Acknowledgements}
We  acknowledge financial
support from the Spanish MEC project FIS2006-04885, the project
CAM-UCM/910758 (AB and MAMD) and the UCM project PR1-A/07-15378 (AL).
Additionally, we acknowledge support from
a FPU MEC grant (AB), and the ESF Science Programme INSTANS 2005-2010 (MAMD).

\vspace{2ex}

\appendix

\section{Standard non-Relativistic limit}
 \label{appendix}

In this appendix we derive the non-relativistic limit of the
two-dimensional Dirac oscillator using the standard techniques in
Relativistic Quantum Mechanics~\cite{greiner}. Let us then
consider equation~\eqref{ec_dirac_oscillator_2D}, where the
relativistic spinor can be rewritten as $|\Psi\rangle:= [ | \psi_1
\rangle ,| \psi_2 \rangle ]^te^{-\ii Et/\hbar}$. Then,
equation~\eqref{ec_dirac_oscillator_2D} becomes a set of coupled
equations
\begin{equation}
\label{ec_do_components}
\begin{array}{c}
  (E-mc^2)| \psi_1 \rangle = c \left[(p_x+\ii m\omega x)-\ii(p_y+\ii m\omega y)\right] |\psi_2 \rangle ,\\
  (E+mc^2) \ket{\psi_2} =c\left[(p_x-\ii m\omega x)+\ii(p_y-\ii m\omega y)\right] \ket{\psi_1}.
\end{array}
\end{equation}
In order to obtain the non-relativistic limit, we must derive the
associated Klein-Gordon equations,  which by virtue of the
canonical commutation relations $[x_j,p_k]=\ii\hbar\delta_{jk}$,
and using the definition of the orbital angular momentum operator
$L_z:=xp_y-yp_x$, become
\begin{equation}
\label{Klein_gordon}
\begin{split}
(E^2-m^2c^4)\ket{\psi_1}=2mc^2\left[H_{\text{ho}}^{\text{2D}}-\hbar\omega-\omega L_z\right]\ket{\psi_1},\\
(E^2-m^2c^4)\ket{\psi_2}=2mc^2\left[H_{\text{ho}}^{\text{2D}}+\hbar\omega-\omega
L_z\right]\ket{\psi_2},
\end{split}
\end{equation}
where we immediately identify the two-dimensional isotropic
harmonic oscillator  $
H_{\text{ho}}^{\text{2D}}=\textbf{p}^2/2m+m\omega^2\textbf{r}^2/2$.
This fact  already shows the connection between the relativistic
Dirac oscillator in Eq.~\eqref{ec_dirac_oscillator_2D} and the
usual non-relativistic harmonic oscillator. The non-relativistic
regime is attained when the relevant energies in the system are
negligible in comparison with the rest mass energy. For the
$\ket{\psi_1}$ component, we let $E=mc^2+\epsilon$ where
$\epsilon\ll mc^2$, so that
\begin{equation}
\label{NR_approx_energy_psi_1}
(E^2-m^2c^4)\approx2mc^2\epsilon+\mathcal{O}\left(\epsilon^2\right).
\end{equation}
Substituting Eq.~\eqref{NR_approx_energy_psi_1} in the
Klein-Gordon equation~\eqref{Klein_gordon}, we obtain the
corresponding non-relativistic limit of the two-dimensional Dirac
oscillator
\begin{equation}
\epsilon\ket{\psi_1}\approx\left[H_{\text{ho}}^{\text{2D}}-\hbar\omega-\omega
L_z\right]\ket{\psi_1}.
\end{equation}
Finally, recovering the total energy $E=mc^2+\epsilon$, we obtain
the following effective Hamiltonian for the non-relativistic limit
\begin{equation}
\label{NR_hamiltonian_psi_1}
H_{\text{eff}}^{(\uparrow)}=mc^2+\left[H_{\text{ho}}^{\text{2D}}-\hbar\omega-\omega
L_z\right],
\end{equation}
where the two-dimensional harmonic oscillator appears together
with the orbital angular momentum. This procedure  must also be applied
to the lower component $\ket{\psi_2}$, where the non-relativistic
limit is attained setting
$E=-mc^2+\epsilon$ where $\epsilon\ll mc^2$
\begin{equation}
\label{NR_approx_energy_psi_2}
(E^2-m^2c^4)\approx-2mc^2\epsilon+\mathcal{O}\left(\epsilon^2\right).
\end{equation}
Substituting once more Eq.~\eqref{NR_approx_energy_psi_2} in the
corresponding Klein-Gordon equation~\eqref{Klein_gordon}, we
obtain
\begin{equation}
\epsilon\ket{\psi_2}\approx-\left[H_{\text{ho}}^{\text{2D}}+\hbar\omega-\omega
L_z\right]\ket{\psi_2},
\end{equation}
which directly leads to the effective Hamiltonian in the
non-relativistic limit by restoring the original energy
$E=-mc^2+\epsilon$
\begin{equation}
\label{NR_hamiltonian_psi_2}
H_{\text{eff}}^{(\downarrow)}=-mc^2-\left[H_{\text{ho}}^{\text{2D}}+\hbar\omega-\omega
L_z\right],
\end{equation}
where the usual two-dimensional harmonic oscillator arises
naturally in this non-relativistic regime. Finally, using the
chiral operators in Eq.~\eqref{circular_operators}, the
two-dimensional harmonic oscillator can also be expressed in terms
of these operators
\begin{equation}
\label{harm_oscillator_chiral}
H_{\text{ho}}^{\text{2D}}=\hbar\omega\left(a_r^{\dagger}a_r+a_l^{\dagger}a_l+1\right),
\end{equation}
which leads to  the corresponding non-relativistic effective
Hamiltonian
\begin{equation}
\label{NR_hamiltonian_chiral}
H_{\text{eff}}=\left[%
\begin{array}{cc}
  mc^2+2\hbar\omega a_l^{\dagger}a_l & 0 \\
  0 & -mc^2-2\hbar\omega (a_l^{\dagger}a_l+1) \\
\end{array}%
\right].
\end{equation}
By virtue of the commutation relations $[a_l,a_l^{\dagger}]=1$, we
can rewrite Eq.~\eqref{NR_hamiltonian_chiral} as
\begin{equation}
\label{NR_effective_hamiltonian_RQM}
H_{\text{eff}}=\left[%
\begin{array}{cc}
  mc^2+2\hbar\omega a_l^{\dagger}a_l & 0 \\
  0 & -mc^2-2\hbar\omega a_la_l^{\dagger} \\
\end{array}%
\right],
\end{equation}
which coincides with the previous derivation using quantum optical
tools~\eqref{NR_effective_hamiltonian_QO}. In this sense, the
insightful Quantum Optics perspective introduced
in~\cite{bermudez_do_2D} offers a better understanding of the
non-relativistic limit which is condensed in fig.~\ref{couplings}.
In this regime, the spinorial levels can be only coupled through
virtual transitions which is translated into a displacement of the
energies~\eqref{energies_NR}.

We finally want to note that the inclusion of  rest mass energy
terms $\pm mc^2$ in Eq.~\eqref{NR_effective_hamiltonian_RQM} is
necessary if one wants to treat both spinorial components
simultaneously. In this limit, these components are associated to
positive- and negative-energy solutions, and their simultaneous
treatment as considered above is a relativistic effect.

\section{Perturbative series of the non-relativistic limit}
\label{appendixB}

In this appendix we derive the complete perturbative series that
arises naturally in the non-relativistic limit. Remarkably, we can
obtain the aforementioned perturbative expansion to every order
$\mathcal{O}\left((\xi n_l)^{k}\right)$ and give a physical
interpretation of the corresponding terms. It turns out that the
whole perturbative series can be interpreted in terms of dynamical
Stark shift terms introduced in Sect.~\ref{sectionIII}, and
interferometric Ramsey processes as those discussed in
Sect.~\ref{sectionV}. In this sense, a complete description of the
phenomenology in the non-relativistic regime of the Dirac
oscillator can be accomplished by only considering the first two
corrections in Sects.~\ref{sectionIII} and~\ref{sectionV}. In
light of these results, we claim that \emph{Zitterbewegung}-like
processes of the Dirac oscillator to any order can be fully
described with the results of this work.

\begin{figure}[!hbp]

\centering

\includegraphics[width=8.5cm]{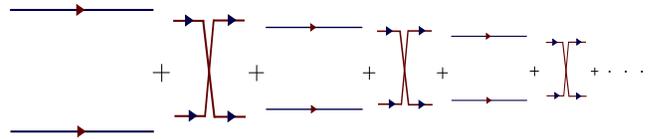}\\

\caption{"(Color
 online)" Diagrammatic scheme of the non-relativistic perturbative
series. The even terms represented by two parallel lines
correspond to the $\sigma_z$ coupling associated to a dynamical
Stark shift term, whereas the odd terms depicted by crossing lines
stand for the $\sigma_y$ interaction associated to a Ramsey
interferometric effect. Note that the different scales stress the
relevance of each term in the non-relativistic regime.
}\label{diagramm_perturbation}

\end{figure}

Let us consider the unitary time evolution operator in
Eq.~\eqref{evolution_operator}, which can be readily expressed as
\begin{equation}
\label{perturbative_te}
U_{n_l}=U^{(0)}+2\ii\sin\phi_{n_l}(t)\sin\theta_{n_l}\left(\sin\theta_{n_l}\sigma_z-\cos\theta_{n_l}\sigma_y\right),
\end{equation}
where $U^{(0)}:=\ee^{-\ii\phi_{n_l}(t)\sigma_z}$ represents the
zeroth order time evolution corresponding to the non-relativistic
limit discussed in Sect.~\ref{sectionIII}. This time evolution can
be interpreted as the dephasing process inside the two arms of the
Mach-Zehnder interferometer ( see fig.~\ref{mach-zender} ). The
remaining term contains the whole perturbative series and
therefore we must expand in powers of the small parameter $\xi
n_l\ll 1$  to obtain the different corrections to the
non-relativistic limit. Considering the following expansions
\begin{equation}
\begin{split}
&\sin^2\theta_{n_l}=\xi n_l-3(\xi n_l)^2+(\xi n_l)^3-35(\xi
n_l)^4+\mathcal{O}\left((\xi n_l)^5\right),
\\
&\sin\theta_{n_l}\cos\theta_{n_l}=(\xi n_l)^{1/2}-2(\xi
n_l)^{3/2}+6(\xi n_l)^{5/2}-\\
&\hspace{14ex}-20(\xi n_l)^{7/2}+\mathcal{O}\left((\xi
n_l)^{9/2}\right),
\end{split}
\end{equation}
the time evolution operator~\eqref{perturbative_te} can be
expressed as follows
\begin{equation}
\label{evolution_operator_pert}
\begin{split}
U_{n_l}&=U^{(0)}+2\ii\sin\phi_{n_l}(t)\left[-(\xi
n_l)^{1/2}\sigma_y +(\xi n_l)\sigma_z+\right.\\
&+2(\xi n_l)^{3/2}\sigma_y-3(\xi n_l)^{2}\sigma_z-6(\xi
n_l)^{5/2}\sigma_y+\\
&\left.+10(\xi n_l)^{3}\sigma_z+20(\xi n_l)^{7/2}\sigma_y -35(\xi
n_l)^{4}\sigma_z\right]+\\
&+\mathcal{O}\left((\xi n_l)^{9/2}\right).
\end{split}
\end{equation}
We observe from this expression how the subsequent perturbative
corrections present either a $\sigma_z$ term associated to a
dispersive Stark-Shift dynamics, or a $\sigma_y$ term which can be
identified with a Ramsey interference effect where the spinorial
components get dynamically mixed during unitary evolution.
Therefore, the perturbative series might be represented
diagrammatically as in fig.~\ref{diagramm_perturbation}.

As we mentioned at the beginning of this appendix, the complete
perturbative series can be interpreted in terms of the dynamical
Stark shift term which accounts for the non-relativistic limit in
section~\ref{sectionIII}, and the Ramsey interference term which
accounts for the first order correction in section~\ref{sectionV}.
Therefore, the full phenomenology and the characterization of
\emph{Zitterbewegung} can be accomplished regarding these two
regimes.

Finally, let us interpret the different terms of the so-called
perturbative expansion with the language of Quantum Optics. The
unitary operator in Eq.~\eqref{evolution_operator_pert} describes
the time evolution inside the invariant subspace
$\mathcal{H}_{n_l}$ in Eq.~\eqref{invariant_subspaces}. If we
recover the full Hilbert space description, we obtain that the
even terms of the perturbative expansion are
\begin{equation}
\label{even_pert_ev_operator}
U^{2k}:=\lambda_{2k}\sin\phi_{n_l}(t)\left[%
\begin{array}{cc}
  (a_l^{\dagger}a_l)^k & 0 \\
  0 & -(a_la_l^{\dagger})^k \\
\end{array}%
\right],
\end{equation}
where $k=0,1...$ and we have introduced certain coupling constants
$\lambda_{2k}(t):=c_{2k}\xi^{k}$ which involve increasing powers
of the relativistic parameter, and $c_{2k}\in\mathbb{C}$ follow
directly from
Eqs.\eqref{perturbative_te},~\eqref{evolution_operator_pert}.
These terms represent a kind of dynamical Stark shift between the
spinorial levels proportional to the $k-$th power of the orbital
quanta number $n_l$. They might be considered as certain shifts
produced by $2k$-virtual transitions between the spinorial levels.

In the same manner, we can rewrite the odd terms as
\begin{equation}
\label{odd_pert_ev_operator}
U^{2k+1}:= \lambda_{2k+1}\sin\phi_{n_l}(t)\left[%
\begin{array}{cc}
  0 &(a_l^{\dagger}a_l)^ka_l^{\dagger}  \\
   -a_l(a_l^{\dagger}a_l)^k & 0 \\
\end{array}%
\right],
\end{equation}
where the coupling parameter
$\lambda_{2k+1}=c_{2k+1}\xi^{\frac{2k+1}{2}}$ becomes more
important as the relarivistic parameter increases, and
$c_{2k+1}\in\mathbb{C}$ also follow from
Eqs.~\eqref{perturbative_te},~\eqref{evolution_operator_pert}.
These perturbative terms~\eqref{odd_pert_ev_operator} can be
directly expresses as a generalized Anti-Jaynes-Cummings evolution
\begin{equation}
\label{AJC_effective_evolution}
U^{2k+1}:=\lambda_{2k+1}\left(\sigma^-A_k-\sigma^+A_k^{\dagger}\right)\sin\phi_{n_l}(t),
\end{equation}
where we have introduced the bosonic operator
\begin{equation}
A_k:=a_l(a_l^{\dagger}a_l)^k.
\end{equation}
 This effective time evolution~\eqref{AJC_effective_evolution}
can be interpreted as a generalized AJC-model where intensity of
the couplings depends on the $k-$th power of the number of
left-handed quanta $n_l$~\cite{knight,vogel}

%%%%%%%%%%%%%%%%%%%%%%%%%%%%%%%%%%%%%%%%%%%%%%%%%%%%%%%%%%%%%%%%%%%%%%%%%%%%%


\begin{references}
\bibitem{comment1}
There are two possible situations where the relativistic system is
mapped onto the usual Jaynes-Cummings
model~\cite{jaynes_cummings}. In an active procedure, the
substitution  $\omega\to\ -\omega$ in
Eq.~\eqref{ec_dirac_oscillator_2D} leads to the chiral partner
Hamiltonian of the Dirac oscillator which can be directly mapped
onto a Jaynes-Cummings interaction~\cite{bermudez_do_2D}.
Conversely, we may regard $\ket{\chi_\uparrow}$ as the ground
state and $\ket{\chi_\downarrow}$ as the excited state, with a
simultaneous change of sign in the detuning $\delta$. This passive
procedure is of no consequence since quantum optical detunings can
experimentally attain both positive and negative values.
\end{references}
\end{document}